\documentclass[12pt,a4paper,reqno]{amsart}
\usepackage{graphicx} 
\usepackage{amsmath,amssymb,amsthm,array}
\usepackage{amsfonts, amsmath, amsthm, amssymb} 

\newcommand{\sech}{\mathop{\mathrm{sech}}\nolimits}

\begin{document}
\title{The KdV soliton crosses a  dissipative and dispersive border}

\author{Alexey Samokhin}\vspace{6pt}

\address{Institute of Control Sciences of Russian Academy of Sciences
65 Profsoyuznaya street, Moscow 117997, Russia}\vspace{6pt}

\email{ samohinalexey@gmail.com}\vspace{6pt}

\begin{abstract}
We demonstrate the behavior of the soliton  which, while moving in non-dissipative and dispersion-constant medium encounters a finite-width barrier with varying  dissipation and/or  dispersion; beyond the layer dispersion is constant (but not necessarily of the same value)  and dissipation is null. The passed wave either retains the form of a soliton (though of different parameters) or becomes a bi-soliton. And a reflection wave may be negligible or absent. This models a situation similar to a light passing from a humid air to a dry one through the vapour saturation/condensation area. Some rough estimations for a prediction of an output are given using relative decay of the KdV conserved quantities are given.
  \vspace{1mm}

\noindent\textbf{Keywords:} KdV- Burgers, non-homogeneous layered media, soliton, bi-soliton, reflection, refraction.

\noindent\textbf{MSC[2010]:} 35Q53, 35B36.
\end{abstract}

\maketitle

\section{Introduction}

 The behavior of solutions to the KdV - Burgers equation is a subject of various recent research, \cite{key-1}--\cite{key-5}. The paper is a continuation of the previous research of the author, \cite{key-5} -- \cite{key-8}, that dealt solely with inhomogeneity of dissipation.

 We demonstrate the behavior of the soliton  which, while moving in non-dissipative and dispersion-constant medium encounters a finite-width barrier with varying  dissipation and/or  dispersion; beyond the layer dispersion is constant (but not necessarily of the same value)  and dissipation is null. The passed wave either retains the form of a soliton (though of different parameters) or becomes a bi-soliton. And a reflection wave may be negligible or absent. This models a situation similar to a light passing from a humid air to a dry one through the vapour saturation/condensation area. Some rough estimations for a prediction of an output are given using relative decay of the KdV conserved quantities are given.

 For the modelling we used the Maple  \emph{PDETools} packet.

 The generalized KdV-Burgers equation considered here is of the form
 \begin{equation}\label{01}
    u_t(x,t)=2u(x,t)u_x(x,t)+g(x)u_{xx}(x,t)+ f(x)u_{xxx}(x,t).
    \end{equation}
 It is the simplest model for the medium which is both viscous and dispersive. The viscosity dampens oscillations except for stationary (or travelling wave) solutions.

 Note that if $f(x)\equiv \mathrm{const}>0$ then for $g(x)\equiv 0$ the equation becomes the KdV equations whose travelling waves solutions are solitons and for $g(x)\equiv\mathrm{const}>0$ becomes the KdV-Burgers equation  whose travelling waves solutions are shock waves.

 In this paper  we consider two possibilities for combinations of $g(x)$ and $f(x)$.

\begin{enumerate}
  \item $g(x)=f'(x)$,  while $f(x)>0$ is a function (numerically) constant outside a finite neighborhood of the origin;
  \item $g(x)=0$ and $f(x)>0$ --- a function which is constant outside a finite neighborhood of the origin.
\end{enumerate}

If $f(\pm\infty)=\gamma_\pm$, then, outside the above mentioned neighborhood, the equation reduces to
$u_t=2uu_x+\gamma_\pm u_{xxx}$. These are the KdV equations whose solitons are of the form $6\gamma_\pm a^2\sech^2(a(x+s)+4\gamma_\pm a^3t)$ and move to the left.

 Hence we use the following initial value --- boundary problem  for the KdV-Burgers equation on $\mathbb{R}$:
 \begin{equation}\label{08}
u(x,0) =6\gamma_+ a^2\sech^2(a(x+s)), \; u(\pm \infty,t) =0,\; u_x(\pm \infty,t) =0.
\end{equation}

Note that the initial datum $u(x,0)$ has a form of the KdV soliton.

 For numerical computations we use $ x\in[a,b]$ for appropriately large $a,\;b$ instead of $\mathbb{R}$.

\section{Dispersion and dissipation; a special case.}

In this section we consider the following equations
\begin{equation}\label{special 1}
  u_t(x,t)=2u(x,t)u_x(x,t)+f'(x)u_{xx}(x,t)+ f(x)u_{xxx}(x,t),
\end{equation}

or

\begin{equation}\label{special 2}
   u_t(x,t)=\left(u_x^2(x,t)+ f(x)u_{xx}(x,t)\right)_x.
\end{equation}

This case models a passage from a half-space with a constant dispersion to a another half-space with different but also constant dispersion; the transition region is dissipative.

Expect each solution  to behave as the one of the KdV at the right half-space  and as a solution of KdV (though with a different coefficient by $u_{xxx}$) at the left one. In our examples we took $f(x)=A+B\tanh(\alpha x)$ or $f(x)=A+B\arctan(\alpha x)$ such that $A\pm B>0$.

The transient wave in a dissipative media transforms to a soliton or a bi-soliton moving to the left; and a reflected wave may be seen in the right half-space. \vspace{3mm}

\subsection{Examples}

\subsubsection{Example 1. Bi-soliton and no reflected wave.}

\begin{figure}[h]
 \begin{minipage}{13.2pc}
\includegraphics[width=13.2pc]{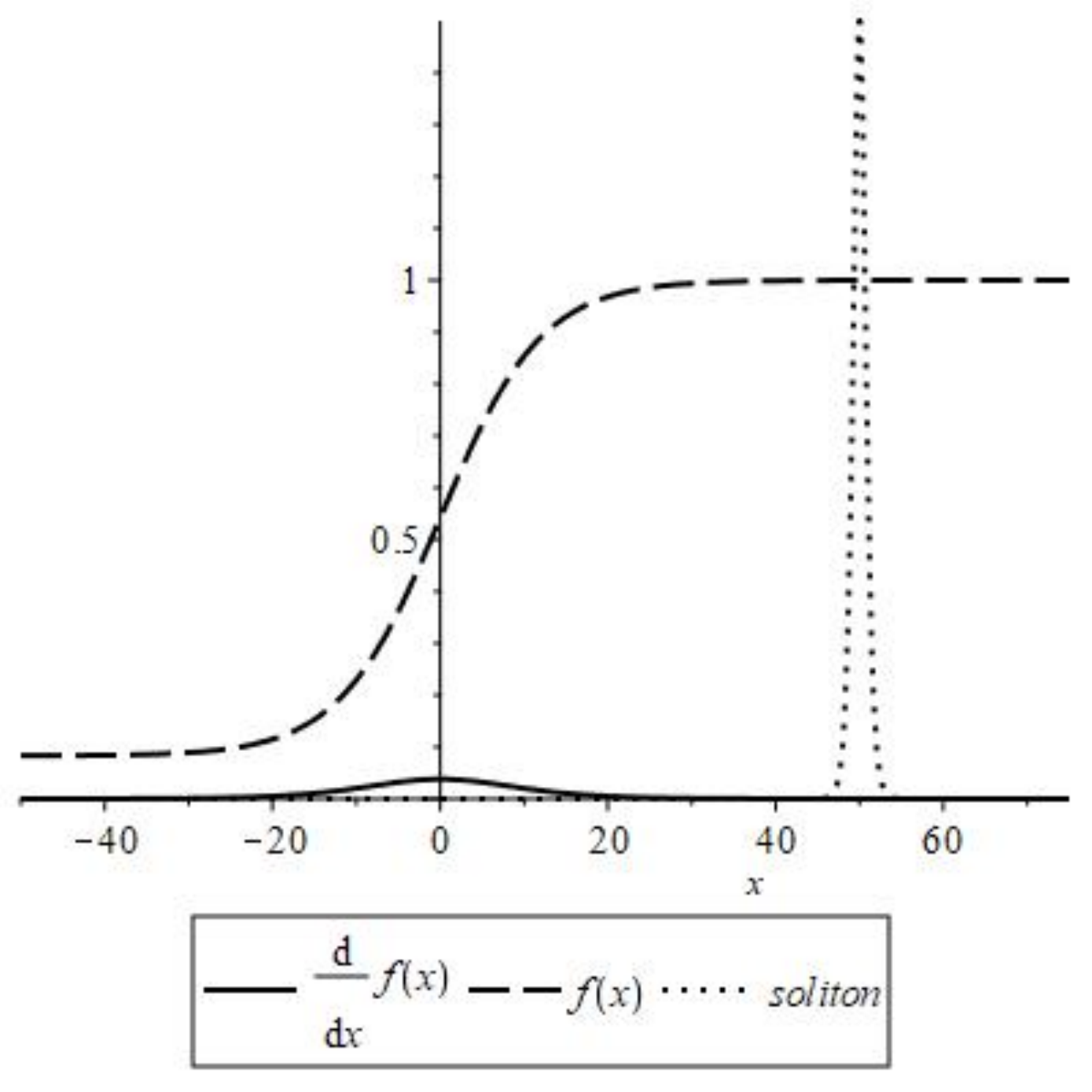}
\end{minipage}
\begin{minipage}{13.2pc}
\includegraphics[width=13.2pc]{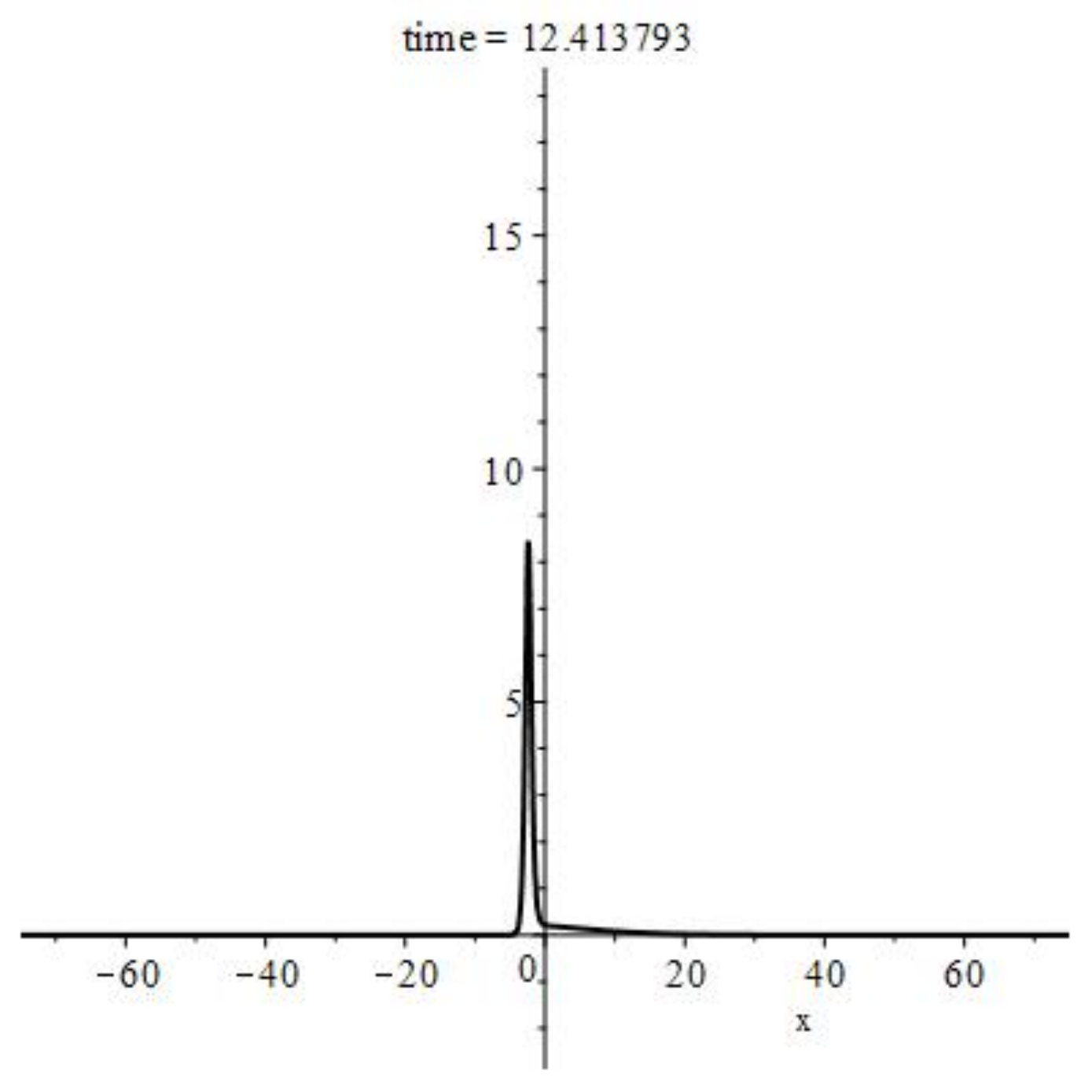}
\end{minipage}
\caption{ \textsl{\textsl{\textbf{Left}:}} Nonhomogeneous layer distribution $f(x)=\frac{1}{24}\left(13+11\tanh(\frac{x}{12})\right)$, $g(x)=f'(x)$ and the initial (scaled-height) soliton $6\sech^2(4t+x-50)$ . 
\textsl{\textbf{Right}:} Soliton $6\sech^2(4t+x-50)$ transformed by the nonhomogeneous layer $\{f(x),g(x)\}$, $t=12$. }
\label{1}
\end{figure}

\begin{figure}[h]
 \begin{minipage}{13.2pc}
\includegraphics[width=13.2pc]{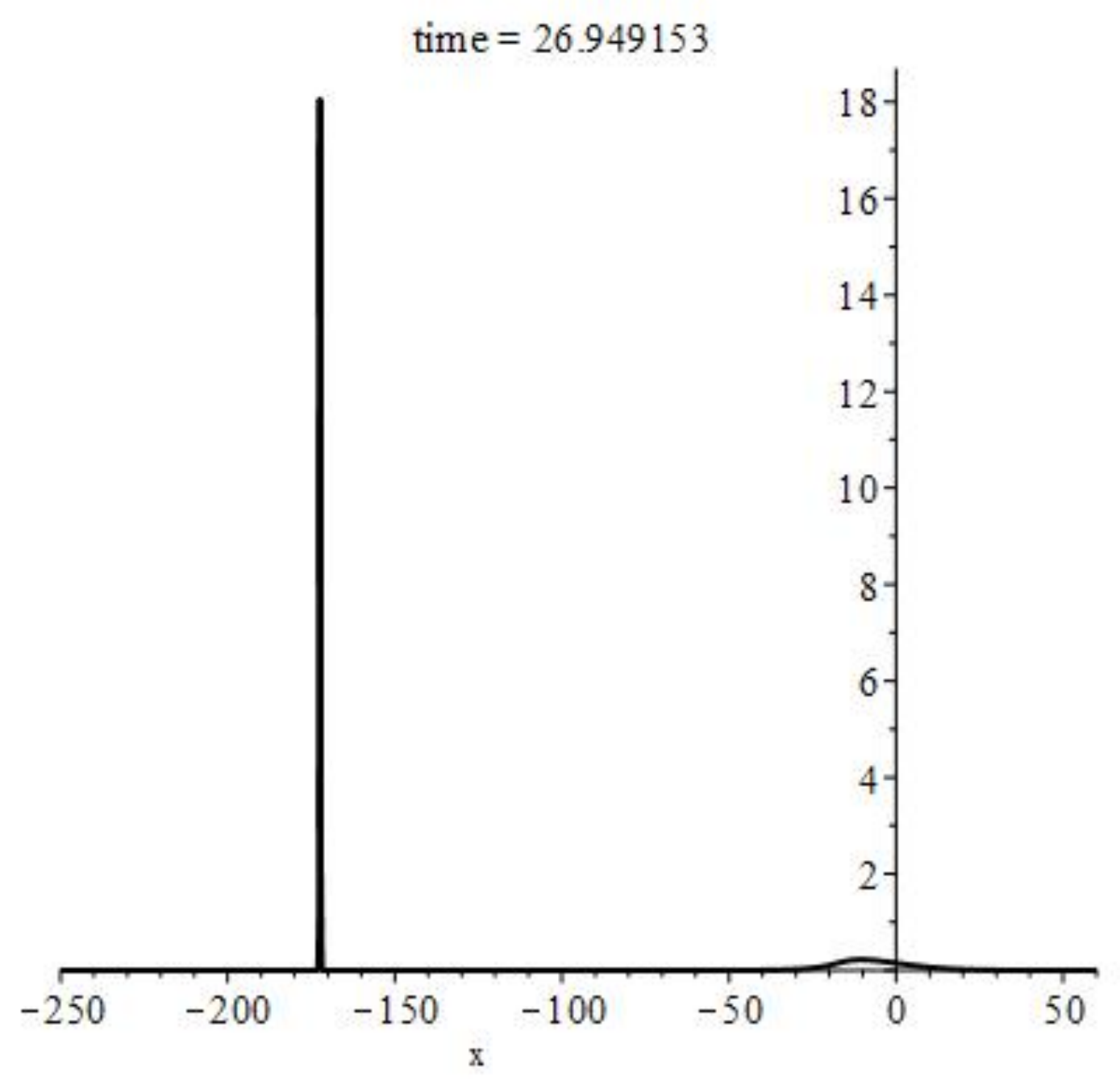}
\end{minipage}
\begin{minipage}{13.2pc}
\includegraphics[width=13.2pc]{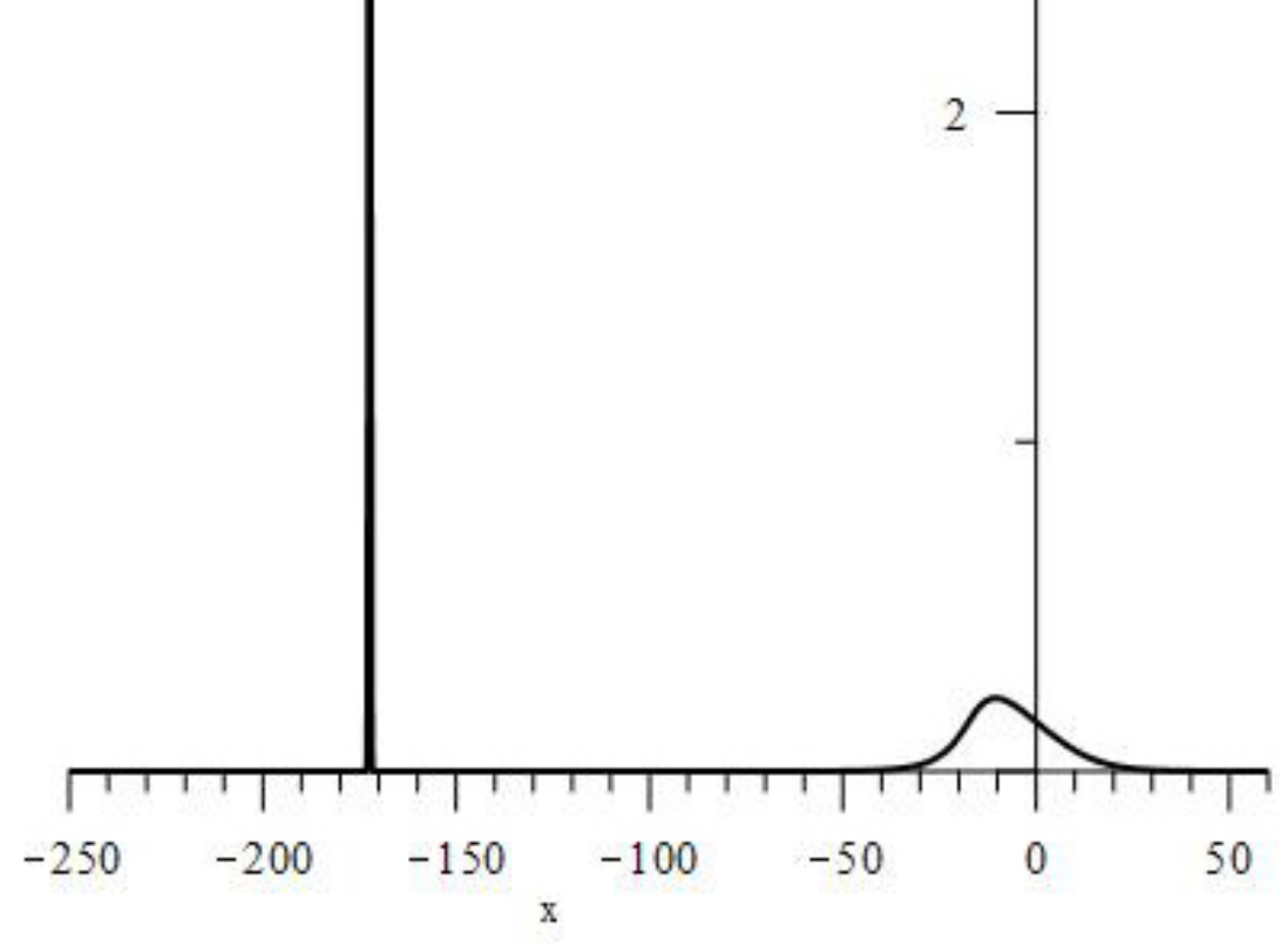}
\end{minipage}
\caption{\textsl{Soliton $6\sech^2(4t+x-50)$  transformed by the nonhomogeneous layer $\{f(x),g(x)\}$, \textbf{Left}: $t=27$.} 
\textsl{\textbf{Right}:}   Enlargement of a part of the previous graph.}
\label{2}
\end{figure}

We chose the decreasing (with respect to the soliton motion) dispersion coefficient $f(x)=\frac{1}{24}\left(13+11\tanh(\frac{x}{12})\right)$ in $u_t=\left(u_x^2+ f(x)u_{xx}\right)_x.$ Thus
$u_t=2uu_x+ u_{xxx}$ at $x=+\infty$ and $u_t=2uu_x+\frac{1}{12} u_{xxx}$ at $x=-\infty$. Results of modeling are presented on figures \ref{1} -- \ref{2}.

No reflected wave can be seen on these graphs.

The stable height of the first peak is about $18.5$. The height of the second one (the peak is under formation, since it have not wholly left the transition region) is about $0.37$. Recall that the amplitude of the initial soliton is $6$. More on this subject below. \vspace{3mm}

\subsubsection{Example 2. Bi-soliton and negligible reflected wave}

\begin{figure}[h]
 \begin{minipage}{13.2pc}
\includegraphics[width=13.2pc]{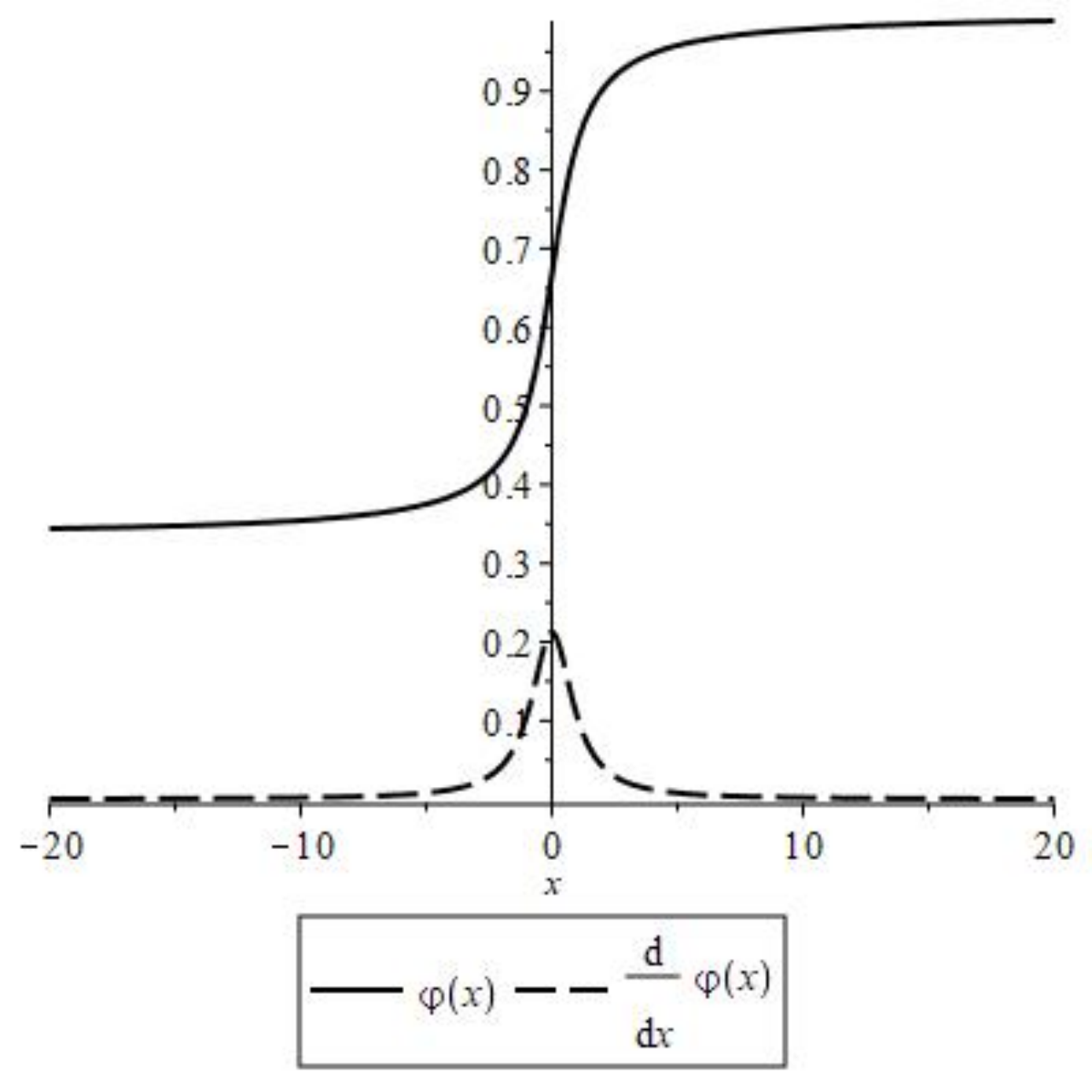}
\end{minipage}
\begin{minipage}{13.2pc}
\includegraphics[width=13.2pc]{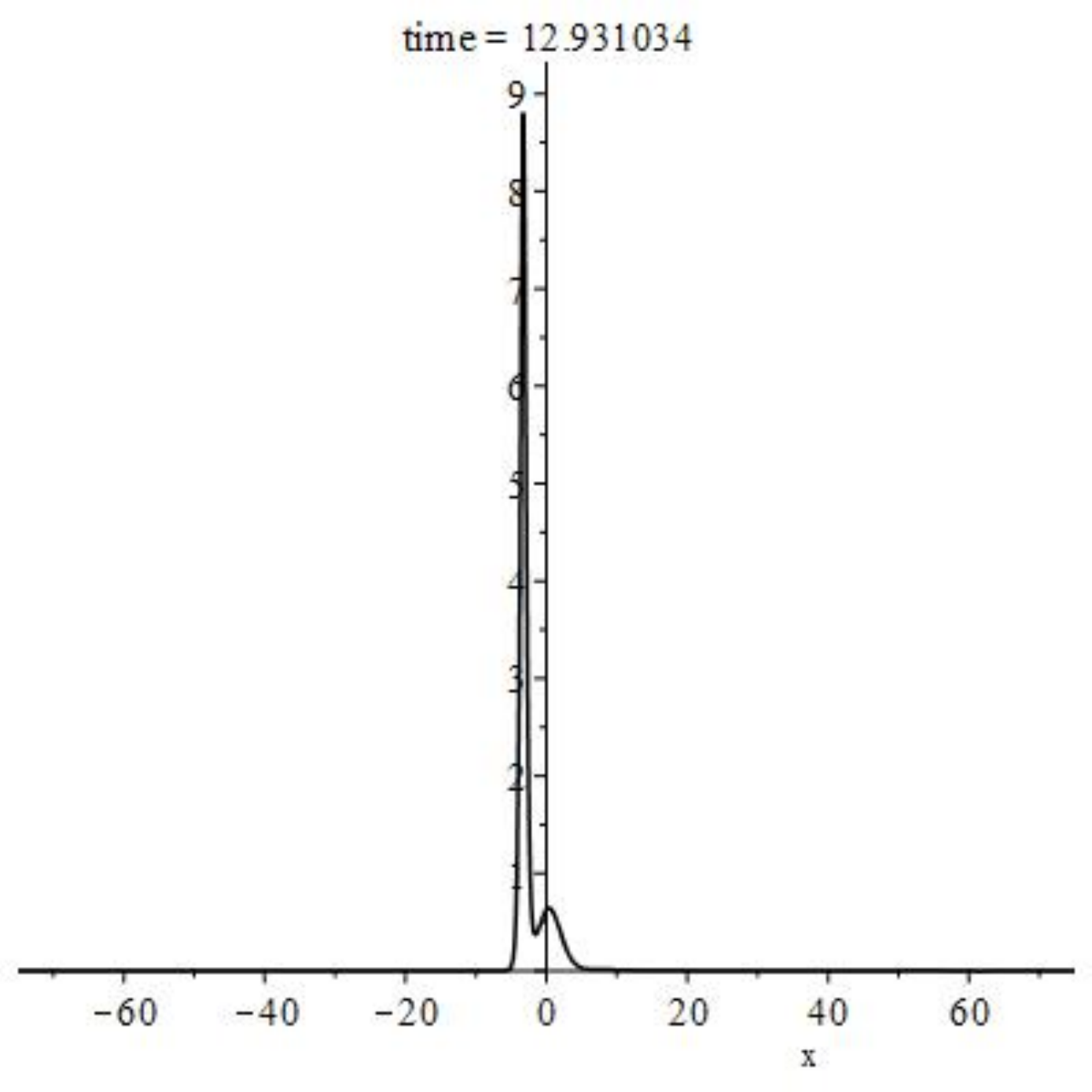}
\end{minipage}
\caption{\textsl{\textbf{Left}:} Nonhomogeneous layer distribution $\varphi(x)=\frac{2}{3}\left(1+\frac{1}{\pi}\arctan(x)\right)$, $\gamma(x)=\varphi'(x)$. 
\textsl{\textbf{Right}:}   $t=13$.}
\label{3}
\end{figure}

\begin{figure}[h]
 \begin{minipage}{13.2pc}
\includegraphics[width=13.2pc]{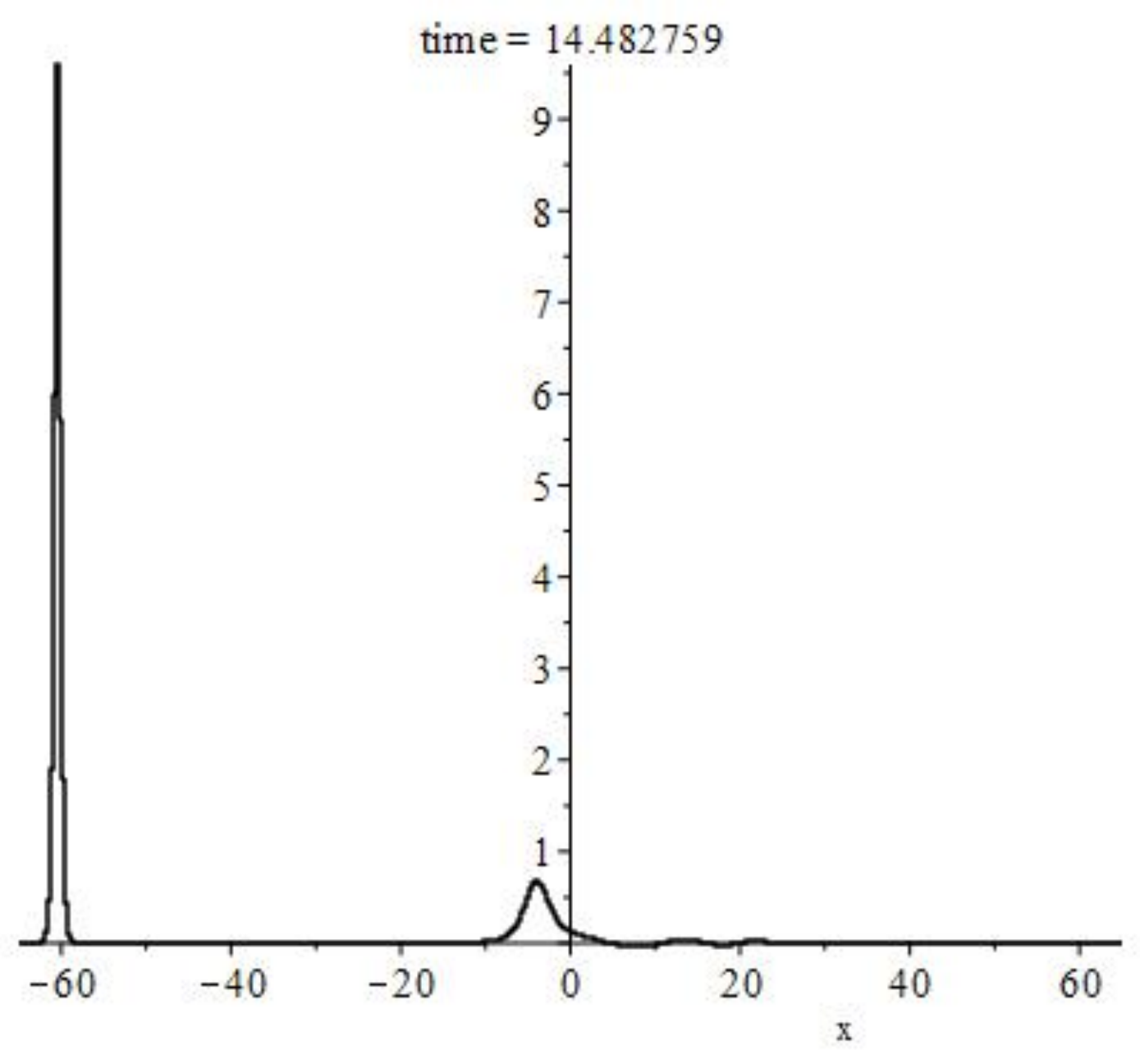}
\end{minipage}
\begin{minipage}{13.2pc}
\includegraphics[width=13.2pc]{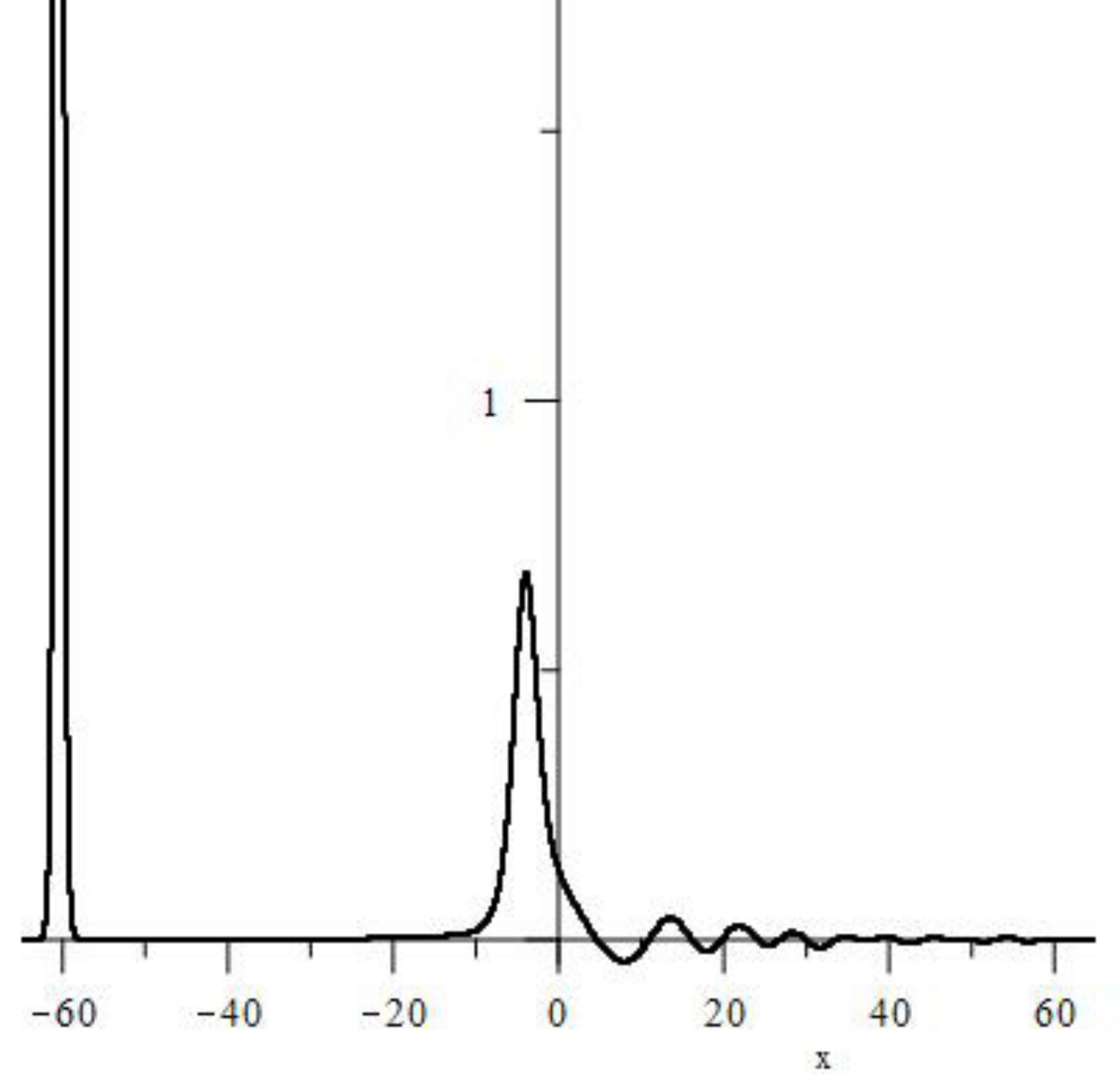}
\end{minipage}
\caption{\textsl{\textbf{Left}:} Soliton $6\sech^2(4t+x-20)$ passing the nonhomogeneous layer $\varphi(x),\varphi'(x)$, $t=15$.
\textsl{\textbf{Right}:}   Enlargement of a part of the previous graph.}
\label{4}
\end{figure}

We chose the decreasing dispersion coefficient $\varphi(x)=\frac{2}{3}\left(1+\frac{1}{\pi}\arctan(x)\right)$ in $u_t=\left(u_x^2+ \varphi(x)u_{xx}\right)_x.$ Thus
$u_t=2uu_x+ u_{xxx}$ at $x=+\infty$ and $u_t=2uu_x+\frac{1}{3} u_{xxx}$ at $x=-\infty$. Results of modeling are presented on figures \ref{3} -- \ref{4}.

A comparatively small reflected wave can be seen as it moves to the right.

The stable height of the first peak is about $9.5$. The height of the second one (the peak is under formation, since it have not wholly left the transition region) is about $0.75$. Recall that the amplitude of the initial soliton is $6$. More on this subject below. \vspace{3mm}

\subsubsection{Example 3. Solitary passed wave and comparable reflected wave.}

\begin{figure}[h]
 \begin{minipage}{13.2pc}
\includegraphics[width=13.2pc]{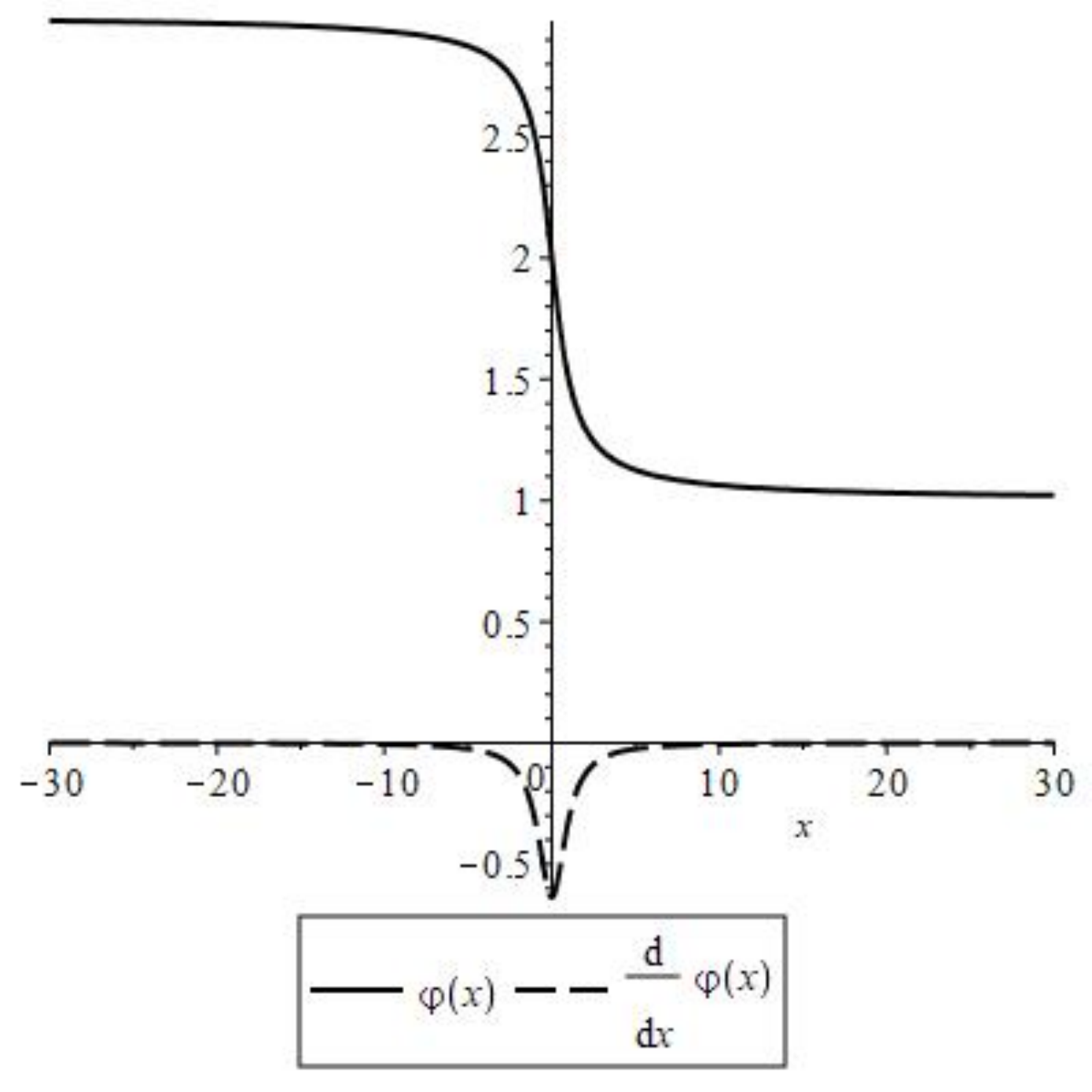}
\end{minipage}
\begin{minipage}{13.2pc}
\includegraphics[width=13.2pc]{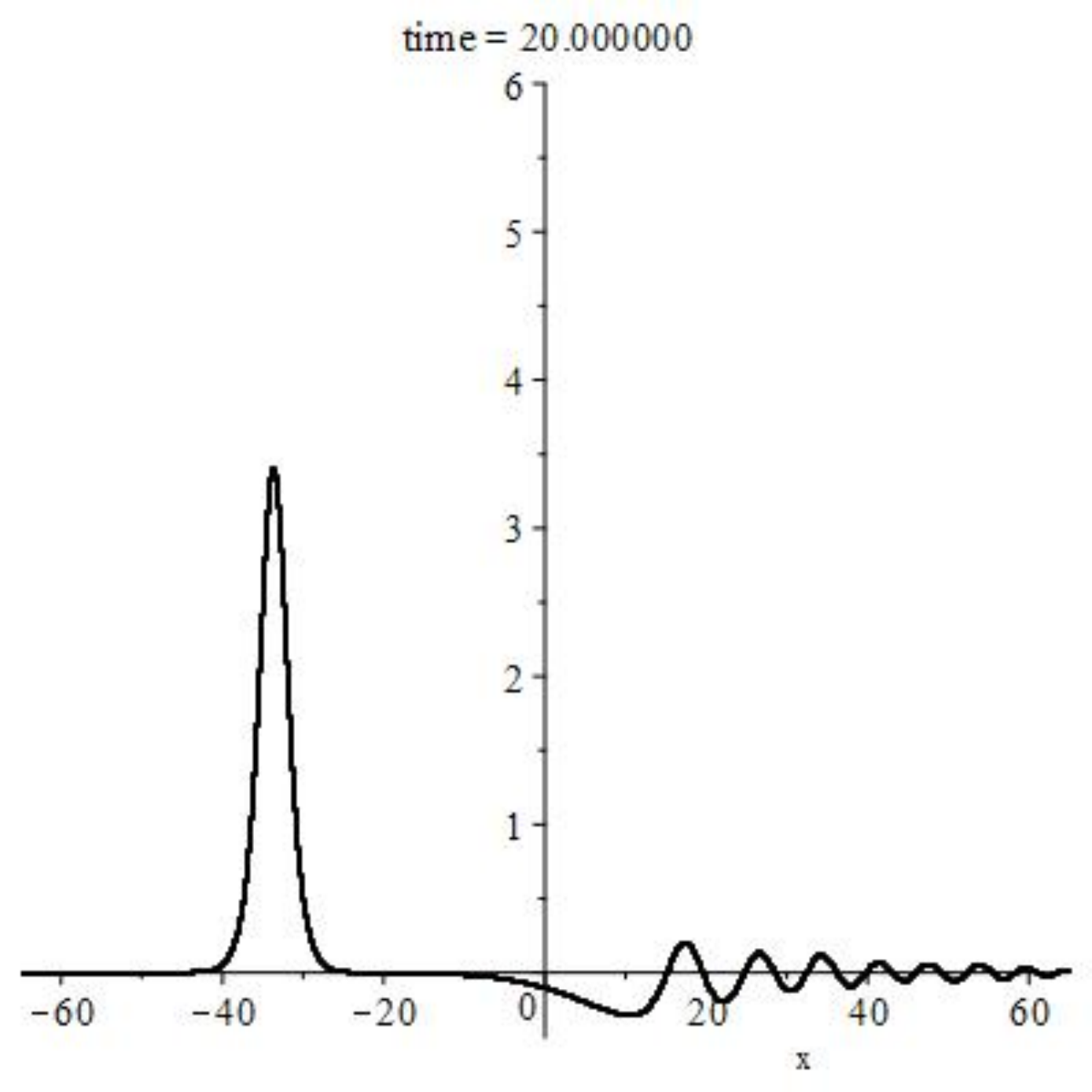}
\end{minipage}
\caption{\textsl{\textbf{Left}:} Nonhomogeneous layer distribution $\varphi(x)=\frac{2}{\pi}\left(\pi-\arctan(x)\right)$, $\gamma(x)=\varphi'(x)$. 
\textsl{\textbf{Right}:} Soliton $6\sech^2(4t+x-20)$ passing the nonhomogeneous layer $\varphi(x),\varphi'(x)$,   $t=20$.}
\label{5}
\end{figure}

We chose the increasing (with respect to the soliton motion)dispersion coefficient $\varphi(x)=\frac{2}{3}\left(1+\frac{1}{\pi}\arctan(x)\right)$ in $u_t=\left(u_x^2+ \varphi(x)u_{xx}\right)_x.$ Thus
$u_t=2uu_x+ u_{xxx}$ at $x=+\infty$ and $u_t=2uu_x+\frac{1}{3} u_{xxx}$ at $x=-\infty$. Results of modeling are presented on figures \ref{5}.

A reflected wave comparable in amplitude with the passed one can be seen.

The stable height of the sole peak is about $3.5$.  Recall that the amplitude of the initial soliton is $6$. More on this subject in the next section.
A comparatively small reflected wave can be seen as it moves to the right.\vspace{3mm}

\subsection{Some \textsl{a priory} estimates\protect\vspace{3mm}}\label{a priory}

\subsubsection{Evolution of the KdV conserved quantities.}

Recall that the soliton $6\gamma a^2\sech^2(a(x+s)+4\gamma a^3t)$ for a KdV equation $u_t=\left(u_x^2+ \gamma u_{xx}\right)_x$ has the amplitude $6\gamma a^2$ and the velocity $4\gamma a^2$.

Since this equation has a form of a conservation law, $u_t=F_x$, the "mass" $\int_{-\infty}^{+\infty}u\,dx$ is a conserved quantity. For a soliton the mass is $12a\gamma$.

In example 1 ($a_0=\gamma_+=1$ for the initial soliton) and there is no reflected wave, so the initial mass $12$ is distributed between two peaks for the $\gamma_-=\frac{1}{12}$ and
\[12a_0=12a_1\frac{1}{12}+12a_2\frac{1}{12},\quad a_1+a_2=12a_0=12.\]
On the other hand, $6a_1^2\frac{1}{12}\approx 18.3$ is the amplitude of the first peak so
\[a_1^2\approx 37,\quad a_1\approx 6.1\]
It follows that $a_2\approx 5.9$ after the second peak leaves the transition region. Its amplitude then will be $17.4$ and velocity $11.6$

By the way, the velocity of the first peak can be measured and it coincides with the theoretical value $4a_1^2\gamma_-\approx 12.4.$

In example 2 one may get a similar if more rough estimations (since it is hard to measure the mass of the reflected wave). In this case $\gamma_+=1,\; \gamma_-=\frac{1}{3}$ and amplitude of the first peak is $9.5.$ So

\[6a_1^2\frac{1}{3}\approx 9.5,\quad a_1\approx 2.18,\quad a_1+a_2=3\Rightarrow a_2\approx 0.82.\]

Consequently, the amplitude and velocity of the second peak are $6a_2^2\gamma_-\approx 1.3$ and $4a_2^2\gamma_-\approx 0.9$ respectively. For the first peak they are approximately $9.5$ and $6.3$.

In example 3 $\gamma_+=1,\; \gamma_-=3$, amplitude is $6a_1^2\cdot 3=3.5$ and there is no second peak. So if $m=\int_0^{+\infty}u(x,t)\,dx,\;t\gg 1$, is the mass of the reflected wave, then
$m=12a_0\gamma_+-12a_1\gamma_-=12-12\cdot 0.44=6.72$

In contrast to the mass, the impulse $\langle u^2\rangle=\int_{-\infty}^{+\infty} u^2\, dx$ is not conserved:

\begin{equation}\label{e2}
  \begin{array}{cc}
    \frac{1}{2}\langle u^2\rangle_t=\langle uu_t\rangle=\langle u(u^2+f(x)u_{xx})_x\rangle & = \\[3mm]
    \langle 2u^2u_x\rangle+\langle u(f(x)u_{xx})_x\rangle & =\\[3mm]
    \left.\frac{2}{3}u^3\right|_{-\infty}^{+\infty} +\left.f(x)uu_{xx}\right|_{-\infty}^{+\infty} -
\langle u_xf(x)u_{xx}\rangle &=  \\[3mm]
    -\frac{1}{2}\left.f(x)u_x^2\right|_{-\infty}^{+\infty}+ \langle f'(x)u^2_{x}\rangle, & \mathrm{so}\\[3mm]
  \langle u^2\rangle_t = 2\langle f'(x)u^2_{x}\rangle & .
  \end{array}
\end{equation}

Thus impulse increases/decreases monotonically whenever  $f'(x)$ is positive/negative (or whenever the dispersion coefficient decreases/increases with respect to the soliton motion). In particular, $f'>0$ in examples 1 and 2; $f'<0$ in examples 3.

For an individual soliton $u=6\gamma a^2\sech^2(a(x+s)+4\gamma a^3t)$ we have  $\langle u^2\rangle=48a^3\gamma^2$.

Thus, in example 2,
$ 48\frac{1}{9}a_1^3+48\frac{1}{9}a_2^3\geqslant 48a_0^3$, i.e. $ a_1^3+a_2^3\geqslant 9$. From the mass conservation law it follows that $a_1+a_2=3$.


The system $\{a_1+a_2=3,\;a_1^3+a_2^3\geqslant 9\}$ implies that the greater parameter $a_1$ satisfies $2<a_1<3$.

Such an additional condition on bi-solitons arises when the system $\{a_1+a_2=\gamma^{-1} a_0,\;a_1^3+a_2^3=\gamma^{-2}a_0\}$ has a solution  $a_1\in (0, \gamma^{-1} a_0)$. In our first example that system has no solutions and $a_1$ may be anywhere in $(\frac{1}{2}\gamma^{-1} a_0),\gamma^{-1} a_0)=(1.5,3)$.\vspace{3mm}

\subsubsection{General case. Refraction coefficient.}

Let $u_t=(u^2+f(x)u_{xx})_x$, $f(+\infty)=\gamma_0, f(-\infty)=\gamma_1$ has a solution $u(x,t)$ such that $u(x,t)=6a_0^2\gamma_0\sech^2(a_0((x+s)+4a^2_0t))|_{t=0}$ and at $t\gg 0$, $u(x,t)$ is a soliton or bi-soliton possibly with reflected wave. Let bi-soliton "consists" of peaks with amplitude $6a_1^2\gamma_1$ and $ 6a_2^2\gamma_1$. If it is plausible to ignore a reflected wave then
\begin{equation}\label{conserv}
  \begin{array}{cc}
    12a_1\gamma_1+ 12a_2\gamma_1= 12a_0\gamma_0 & - \mbox{mass conservation;} \\[3mm]
   48a_1^3\gamma_1^2+ 48a_2^3\gamma_1^2> 48 a_0^3\gamma_0^2 & - \mbox{impulse evolution, } f'\geqslant 0;\\[3mm]
    48a_1^3\gamma_1^2+ 48a_2^3\gamma_1^2< 48 a_0^3\gamma_0^2 & - \mbox{impulse evolution, } f'\leqslant 0.
  \end{array}
\end{equation}

Denote
\[y=\frac{a_1\gamma_1}{a_0\gamma_0},\quad z=\frac{a_2\gamma_1}{a_0\gamma_0}, \quad k=\frac{\gamma_1}{\gamma_0};
 \]
 the \eqref{conserv} may be rewritten to the form

\begin{equation}\label{conserv1}
  \begin{array}{cc}
    y+z=1 & - \mbox{mass conservation;} \\[3mm]
  y^3+z^3>k & - \mbox{impulse evolution, } f'\geqslant 0;\\[3mm]
    y^3+z^3>k & - \mbox{impulse evolution, } f'\leqslant 0.
  \end{array}
\end{equation}

The solution of the system $\{ y+z=1,\;   y^3+z^3=k\}$ is $\{\frac{1}{2}\pm\frac{1}{6}\sqrt{12k-3},\frac{1}{2}\mp\frac{1}{6}\sqrt{12k-3} \}$. Since obviously  $0\leqslant y,\;z\leqslant 1$, it make sense only for $\frac{1}{4}\leqslant k\leqslant 1$, see figure \ref{restr}.

In this case for the first (greater) peak it follows that
 \[ 1>y=\frac{a_1\gamma_1}{a_0\gamma_0}>y_+=\frac{1}{2}+\frac{1}{6}\sqrt{12k-3}.\]

 Since the refraction coefficient
 $R=\frac{V_1}{V_0}=\frac{4a_1^2\gamma_1}{4a_2^0\gamma_0} \mbox{ equals } \frac{y^2}{k}$
 we obtain the restriction on the first peak refraction coefficient (it also coincides with the amplitudes ratio)
 \[ R>\frac{2k+1+\sqrt{12k-3}}{6k}.\]
It is relevant only for  $\frac{1}{4}\leqslant k\leqslant 1, $ see figure \ref{7}, left.

\section{Dispersion, but no dissipation}

\subsection{Examples}

Here we study the evolution of a soliton solution to the equation $u_t=2uu_x+f(x)u_{xxx}$, in non-dissipative media. That is, $g(x)=0$ and $f(x)>0$ --- a function which is constant outside a finite neighborhood of the origin. Below $f(+\infty)=1$ and $u(x,0)=6\sech^2(4t+x-20)|_t=0.$

Only results of mathematical modelling are presented in this section. They give some idea of a range of possibilities in this case.

\subsubsection{Example 4}

Here $f(x)=2-\tanh(x)$, s0 dispersion increases in the path of the soliton.

\begin{figure}[h]
\begin{minipage}{13.2pc}
\includegraphics[width=13.2pc]{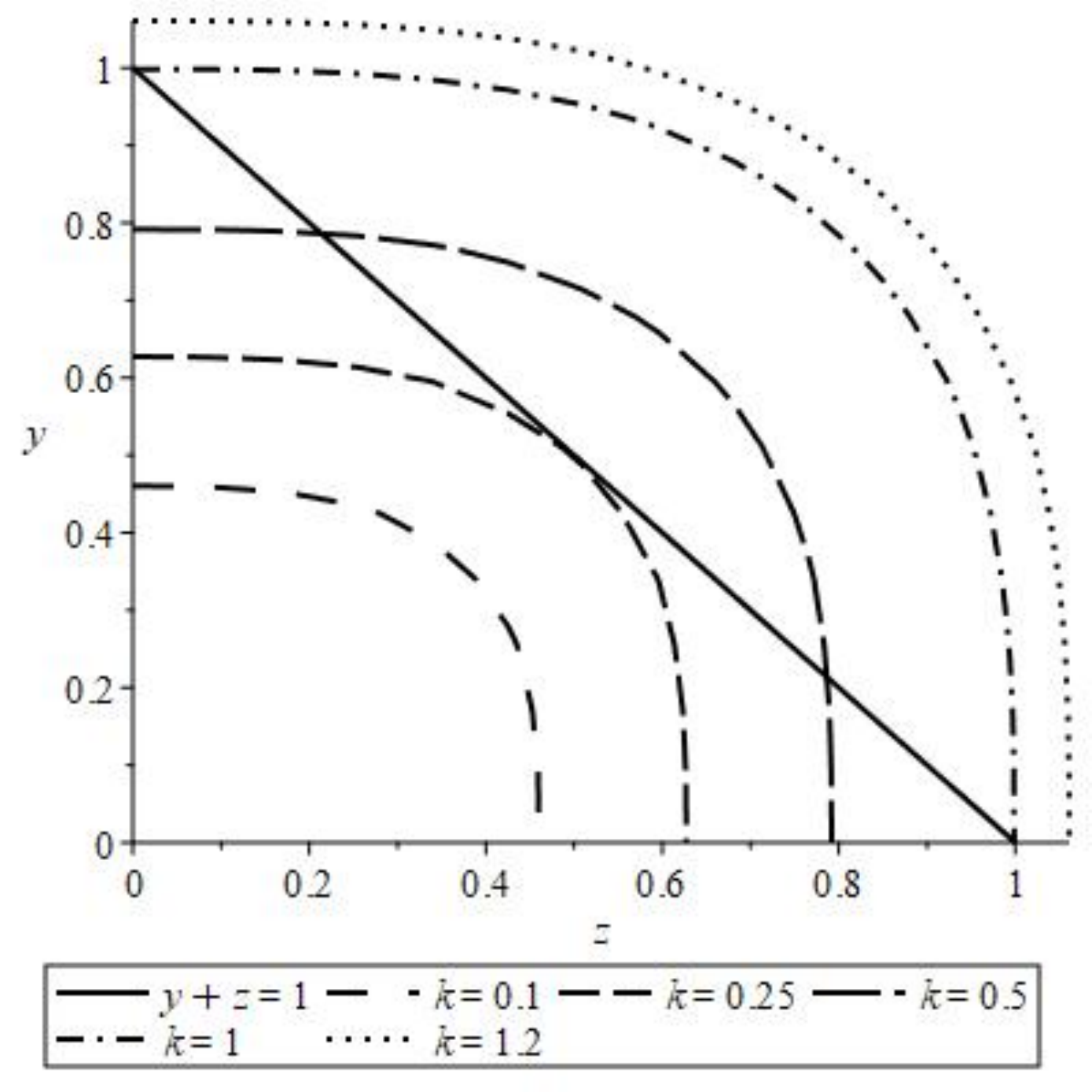}
\end{minipage}
\begin{minipage}{13.2pc}
\includegraphics[width=13.2pc]{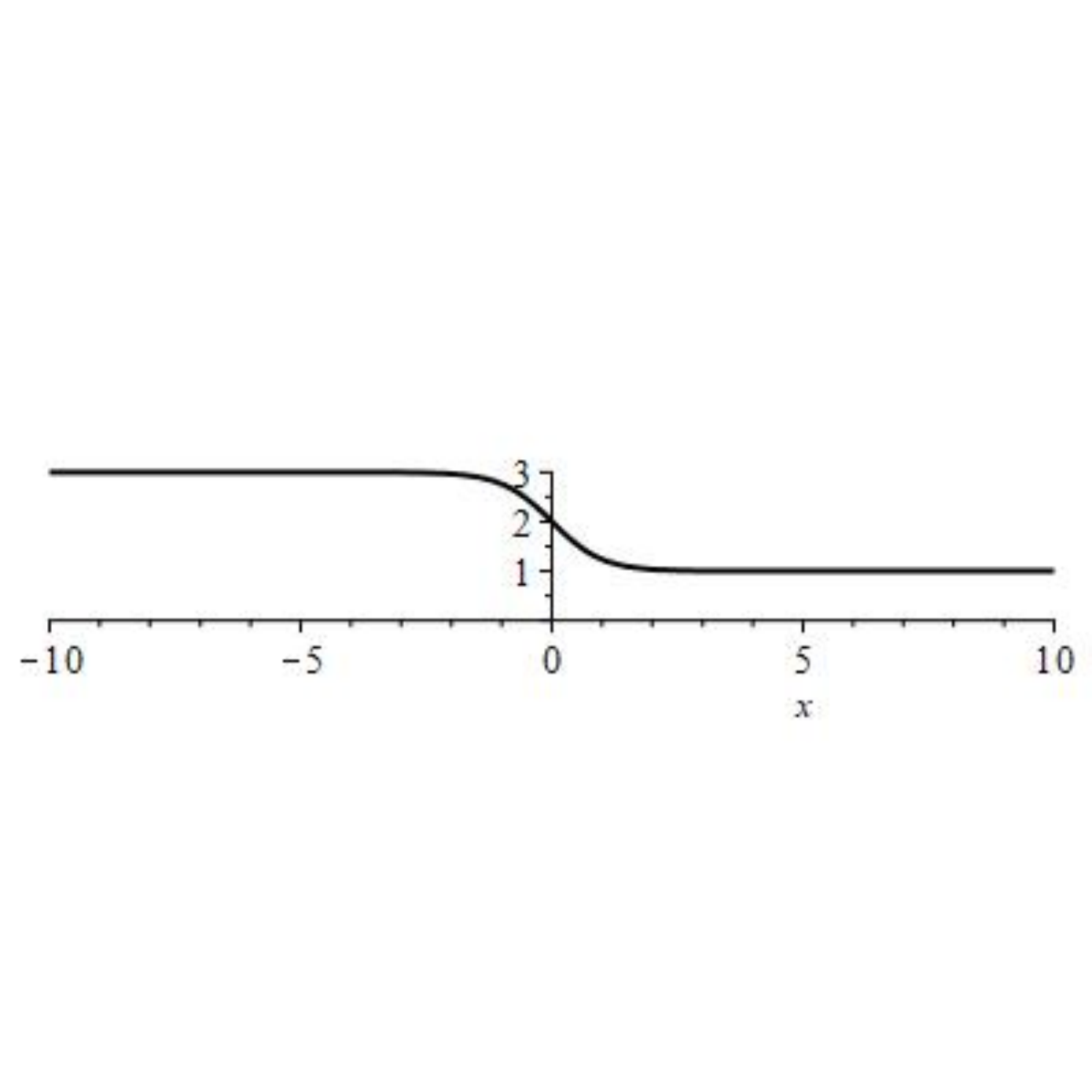}
\end{minipage}
\caption{\protect\small\textsl{\textbf{Left}:} $y+z=1$ and level curves for $y^3+z^3=k$ 
\textsl{\textbf{Right}:} Nonconstant dispersion distribution $f(x)=2-\tanh(x)$.}\label{7}
\end{figure}

\begin{figure}[h]
 \begin{minipage}{13.2pc}
\includegraphics[width=13.2pc]{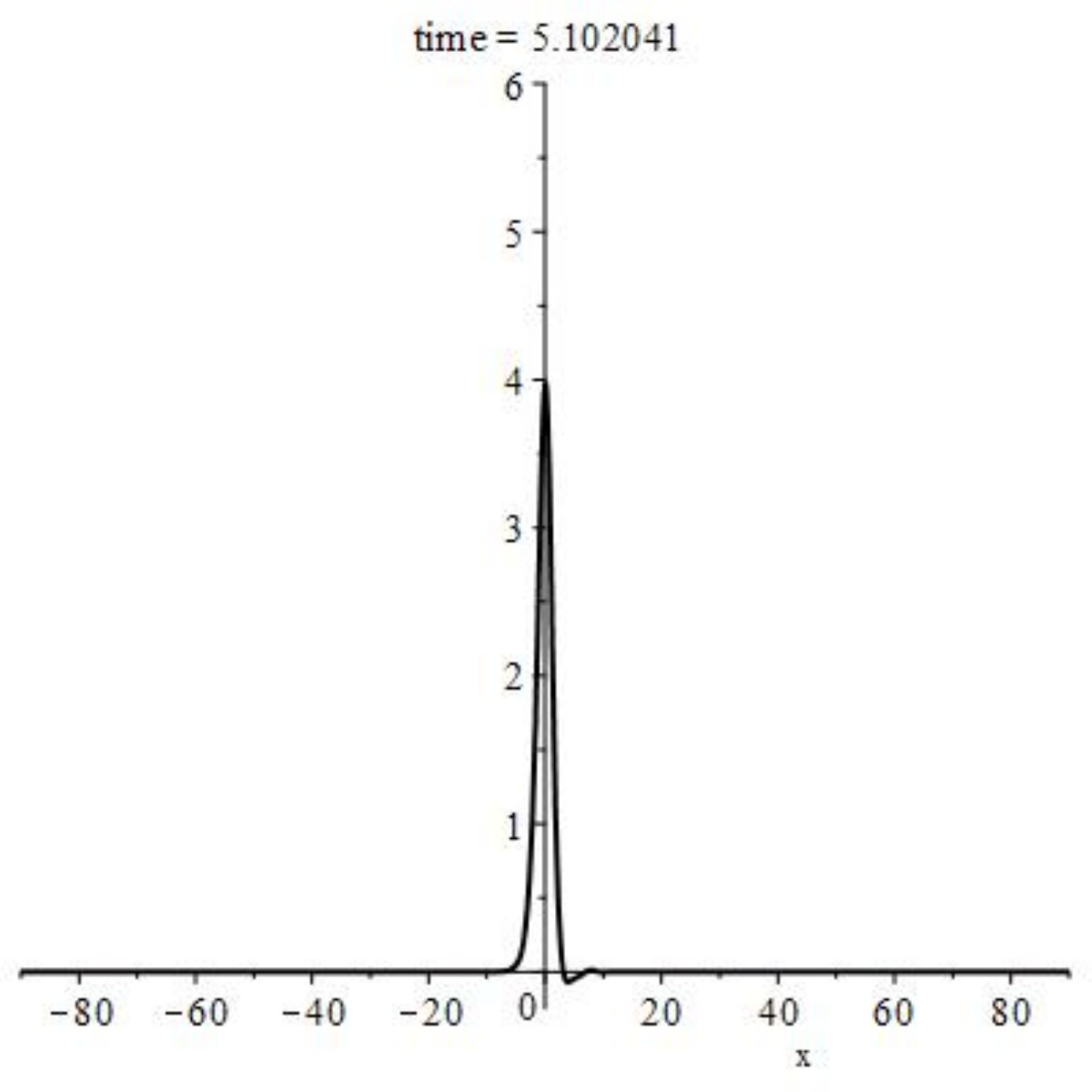}
\end{minipage}
 \begin{minipage}{13.2pc}
\includegraphics[width=13.2pc]{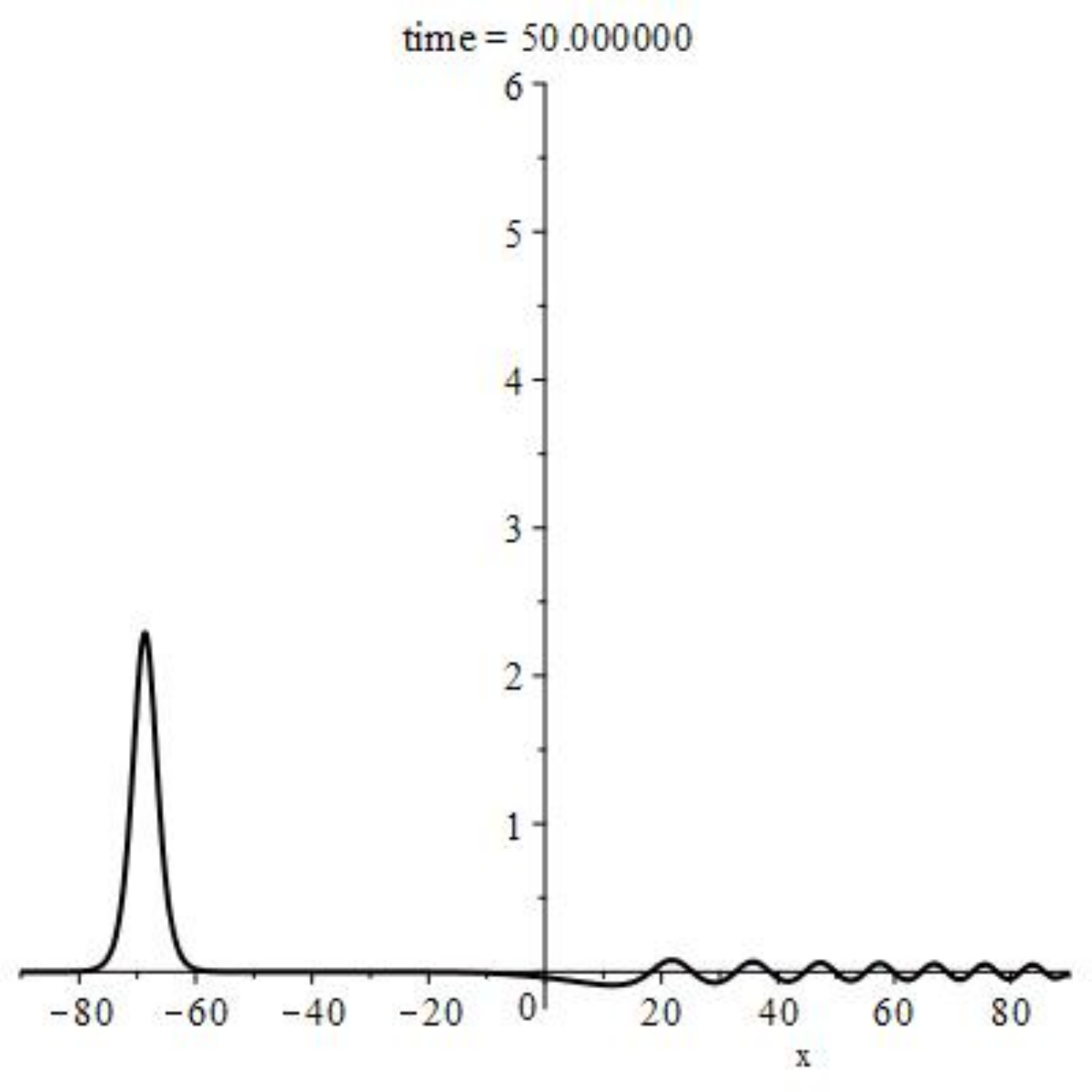}
\end{minipage}
\caption{ Soliton $u(x,t)=6 \sech^2(4t+x-20)$ passing the nonconstant dispersion layer,  dispersion distribution $f(x)=2-\tanh(x)$ \textsl{\textbf{Left}: $t=4$}. 
\textsl{\textbf{Right}:}  $t=50$.}\label{8}
\end{figure}

A singe soliton emerges and reflected wave is comparatively small, see figures \ref{7}--\ref{8}.

\subsubsection{Example 5}

Here nonconstant dispersion layer is given by $f(x)=1-\frac{2}{3}\sech^2(\frac{x}{3})$.

\begin{figure}[h]
  \begin{minipage}{13.2pc}
\includegraphics[width=13.2pc]{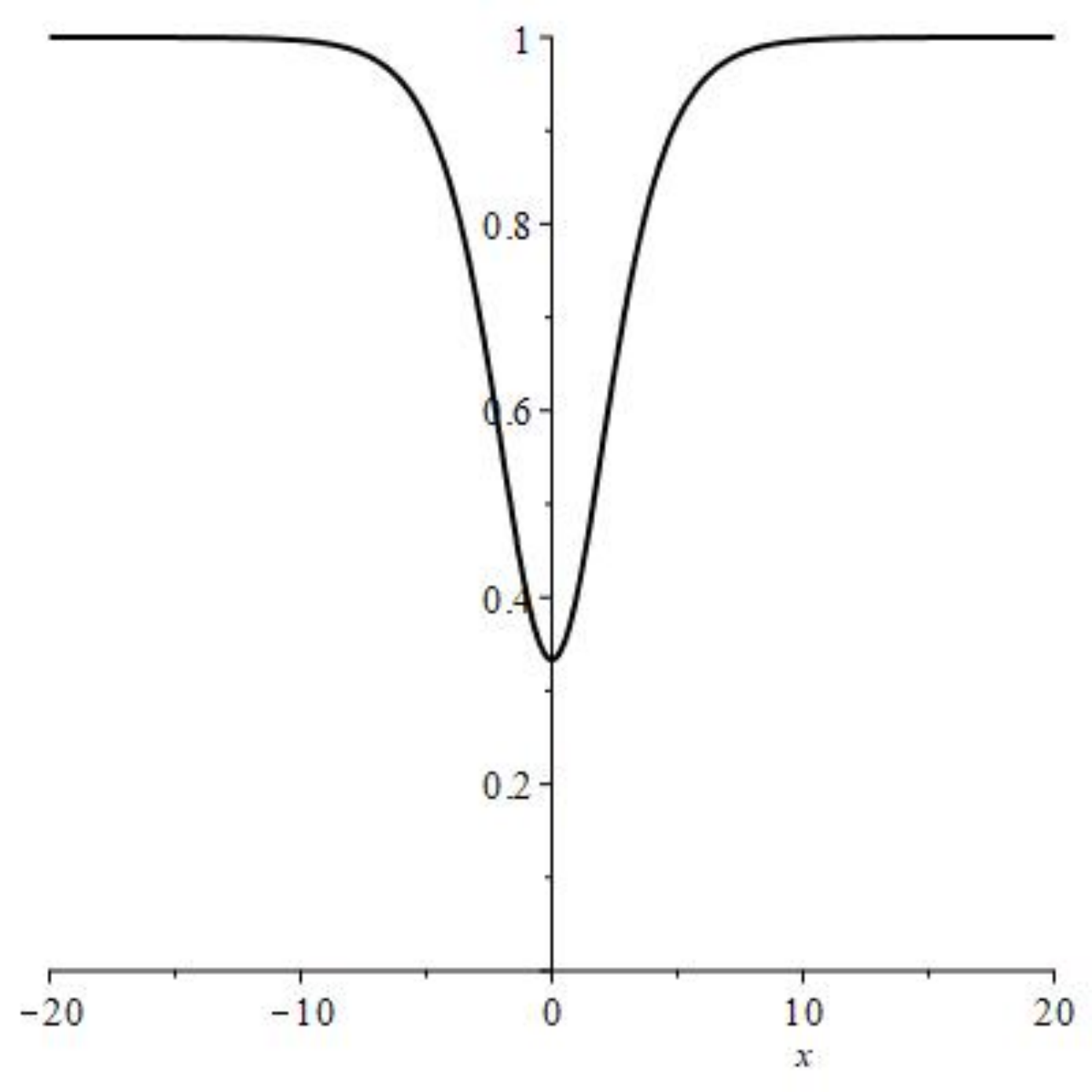}
\end{minipage}
 \begin{minipage}{13.2pc}
\includegraphics[width=13.2pc]{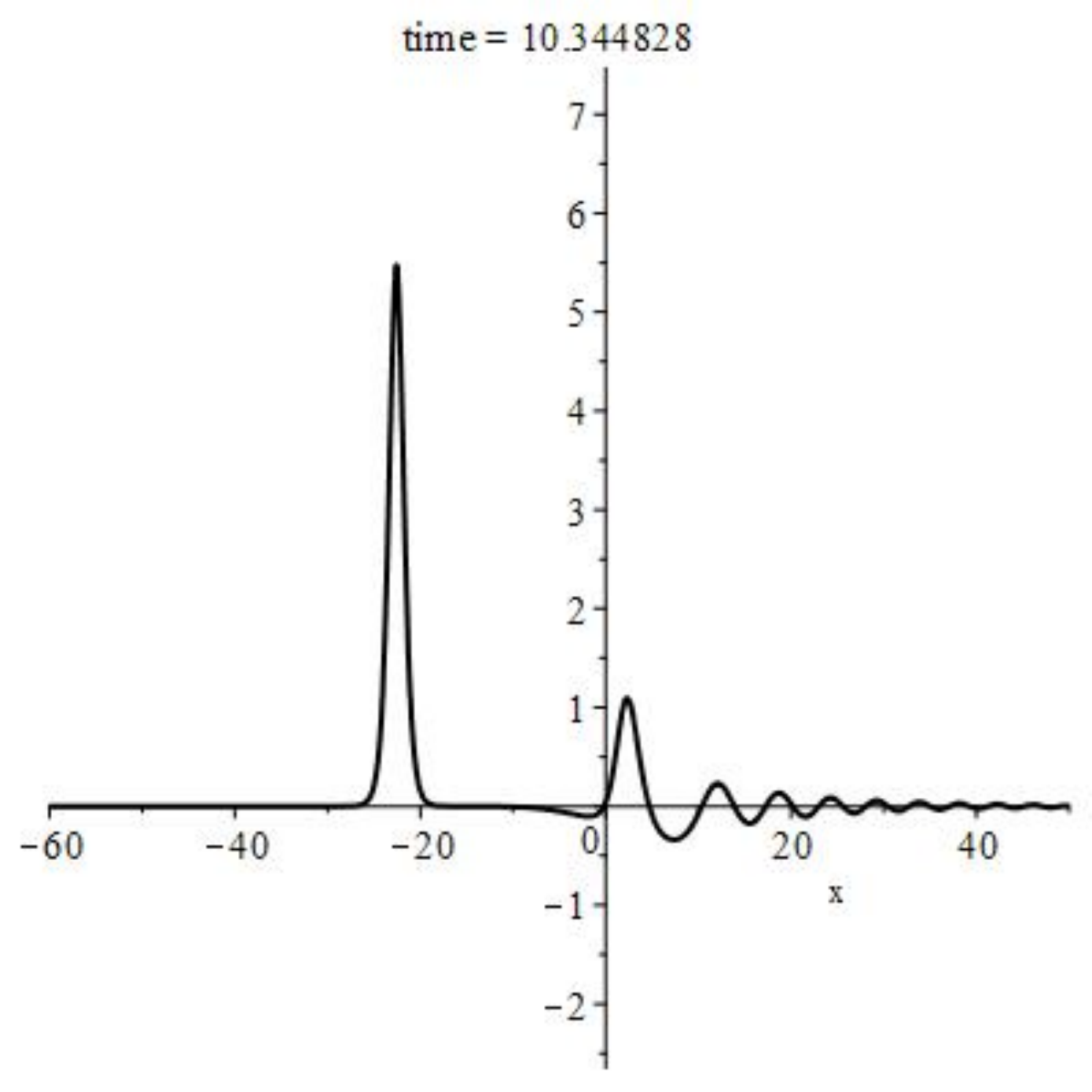}
\end{minipage}
\caption{\textsl{\textbf{Left}:} Nonconstant dispersion layer ($f(x)=1-\frac{2}{3}\sech^2(\frac{x}{3})$). 
\textsl{\textbf{Right}:} Soliton $6\sech^2(4t+x-20)$ passing the  nonconstant dispersion layer $f(x)=1-\frac{2}{3}\sech^2(\frac{x}{3})$, $t=5$.}\label{9}
\end{figure}

\begin{figure}[h]
\begin{minipage}{13.2pc}
\includegraphics[width=13.2pc]{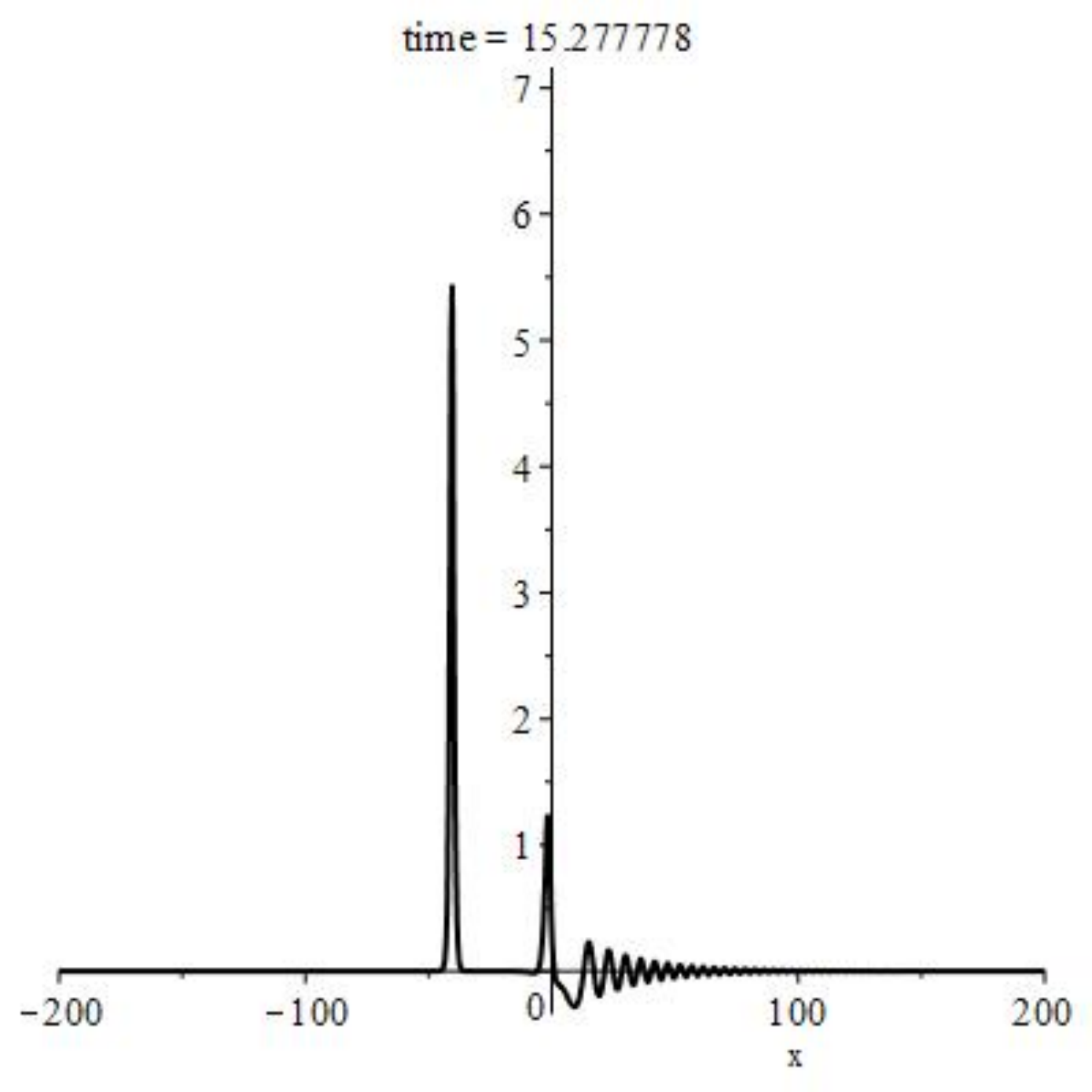}
\end{minipage}
\begin{minipage}{13.2pc}
\includegraphics[width=13.2pc]{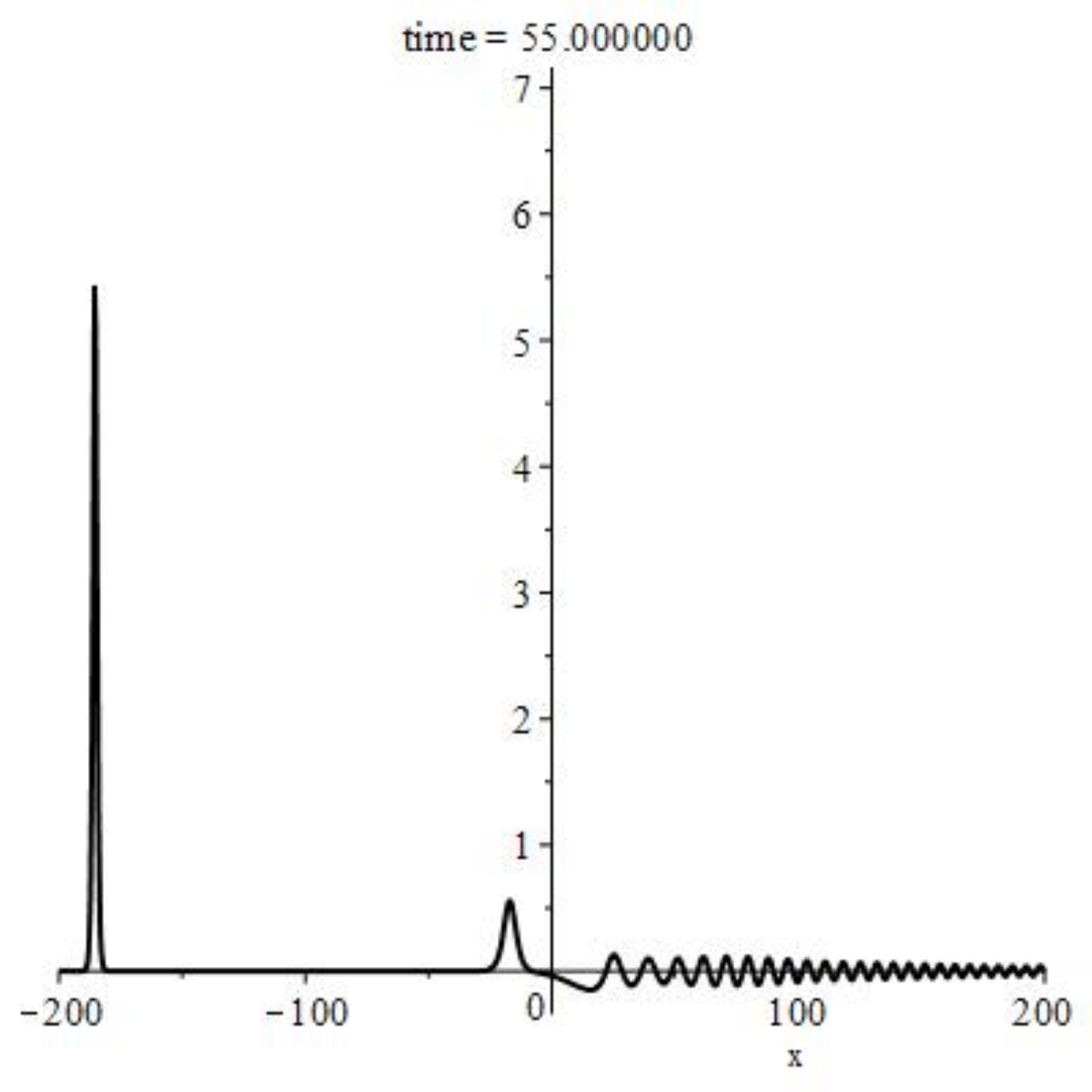}
\end{minipage}
\caption{\textsl{\textbf{Left}:} Soliton $6\sech^2(4t+x-20)$ passing the  nonconstant dispersion layer $f(x)=1-\frac{2}{3}\sech^2(\frac{x}{3})$, $t=15$. 
\textsl{\textbf{Right}:}   $t=55$}.\label{10}
\end{figure}

A bi-soliton develops and reflected wave is comparatively small, see figures \ref{9}--\ref{10}.\newpage

\subsubsection{Example 6}

In this case the nonconstant dispersion layer is $f(x)=1+\frac{2}{3}\sech^2(\frac{x}{3})$. This time the distribution has a peak, contrary to the previous example where the curve dips.

\begin{figure}[h]
\begin{minipage}{13.2pc}
\includegraphics[width=13.2pc]{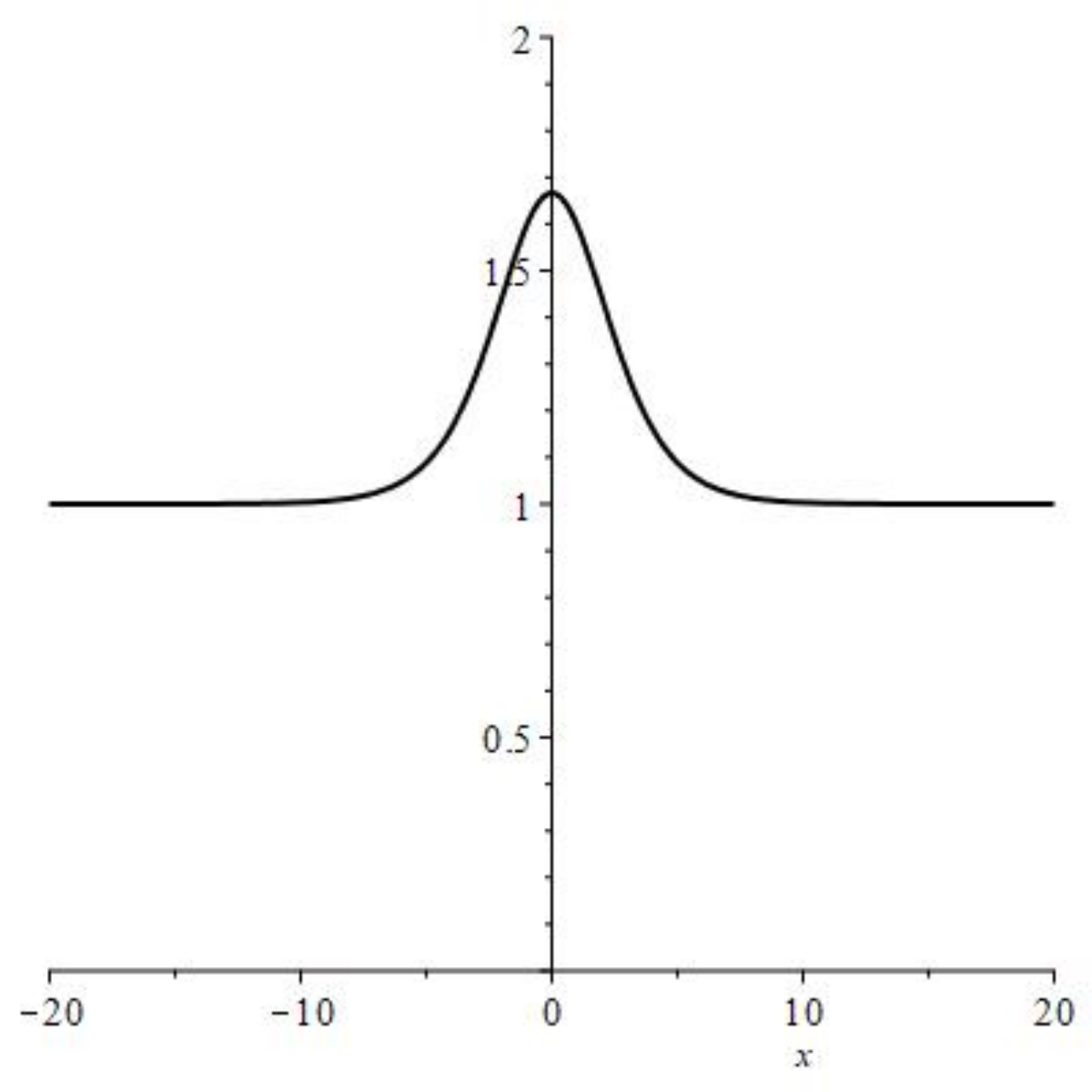}
\end{minipage}
\begin{minipage}{13.2pc}
\includegraphics[width=13.2pc]{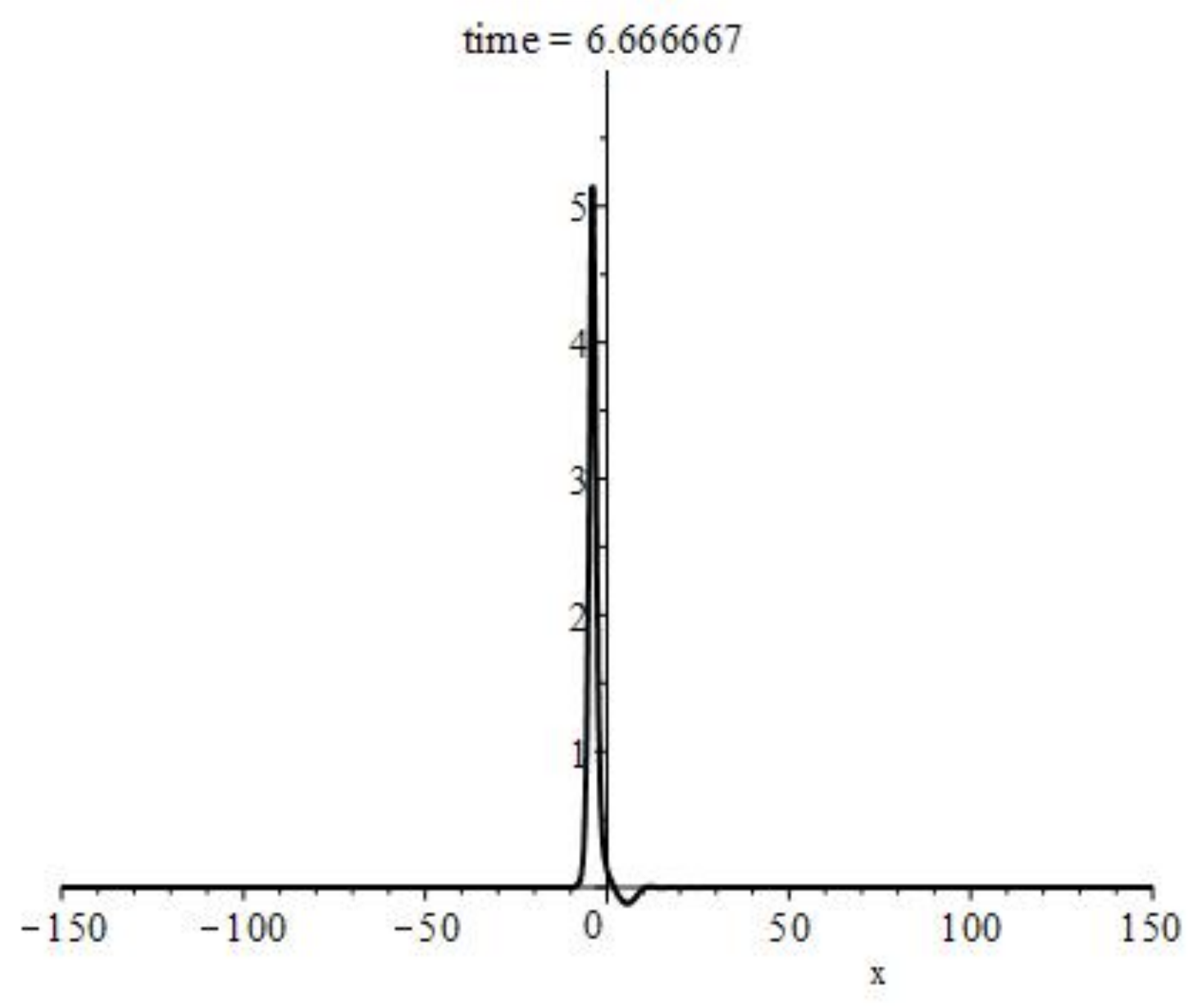}
\end{minipage}\caption{\textsl{\textbf{Left}:} Nonconstant dispersion layer ($f(x)=1+\frac{2}{3}\sech^2(\frac{x}{3})$). 
\textsl{\textbf{Right}:} Soliton $6\sech^2(4t+x-20)$ passing the  nonconstant dispersion layer $f(x)=1+\frac{2}{3}\sech^2(\frac{x}{3})$, $t=7$. }\label{11}
\end{figure}

\begin{figure}[h]
\begin{minipage}{13.2pc}
\includegraphics[width=13.2pc]{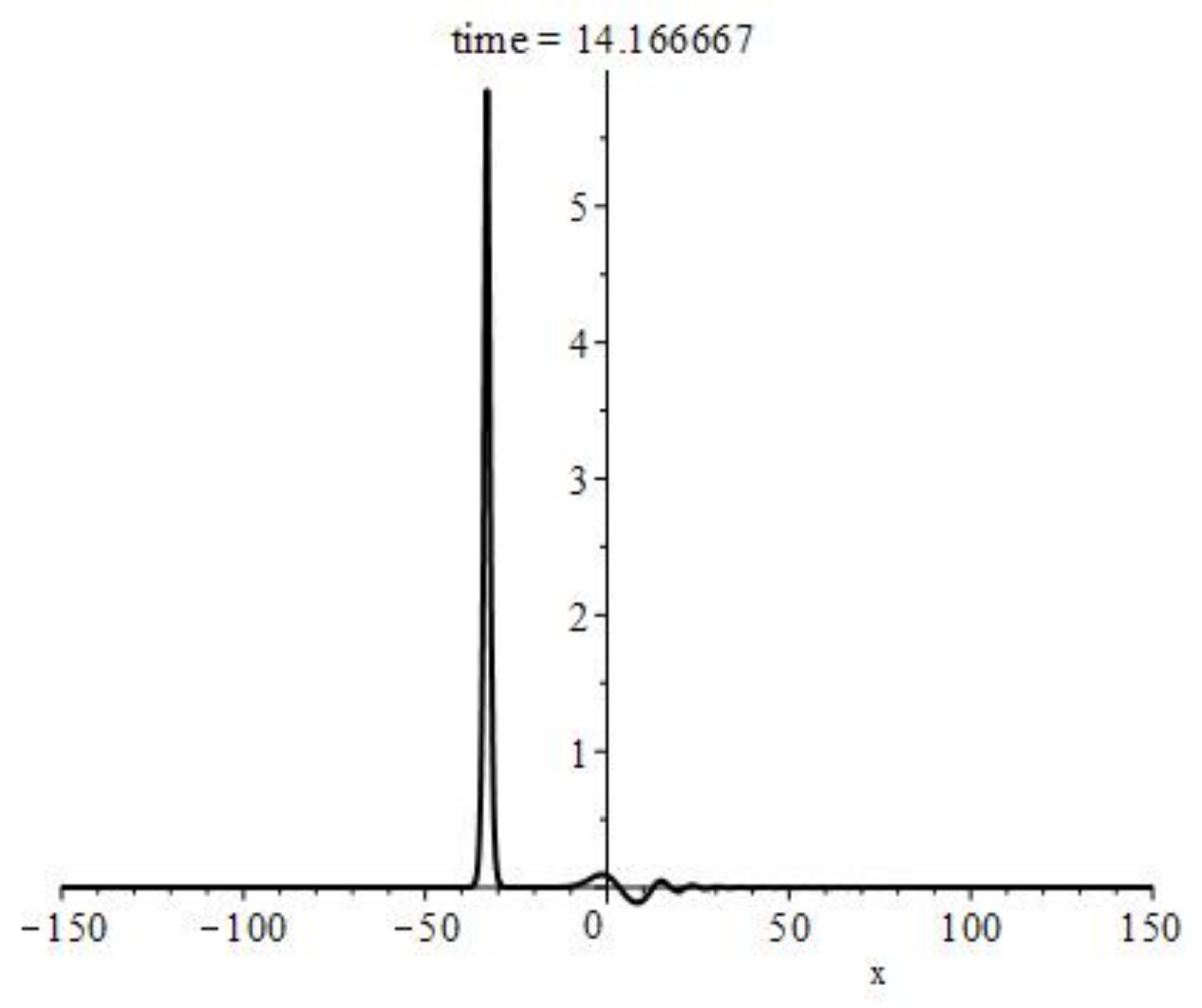}
\end{minipage}
\begin{minipage}{13.2pc}
\includegraphics[width=13.2pc]{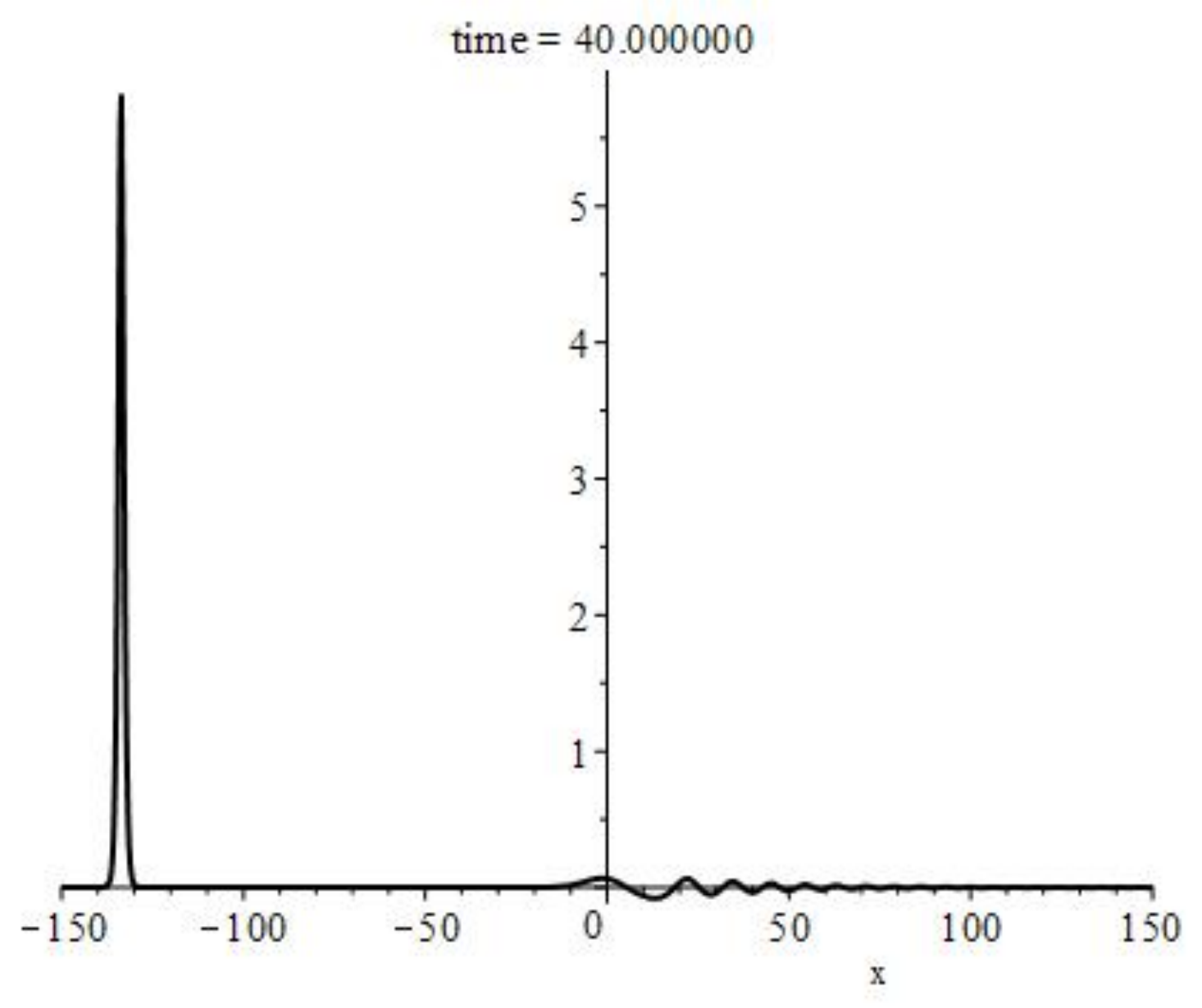}
\end{minipage}\caption{\textsl{\textbf{Left}:} Soliton $6\sech^2(4t+x-20)$ passing the  nonconstant dispersion layer $f(x)=1+\frac{2}{3}\sech^2(\frac{x}{3})$, $t=14$. 
\textsl{\textbf{Right}:}   $t=40$}.\label{12}
\end{figure}

A single soliton develops and reflected wave is  small, see figures \ref{11}--\ref{12}.

\subsection{Discussion}

The present paper as well as our previous research of the KdV solitons in inhomogeneous media (\cite{key-4,key-5,key-6,key-7}) persuades that a distorted by inhomogeneity compact impulse getting into homogeneous region behaves  according the same scenario: it became a soliton or scatter into two or more  of them. Usually, but not necessarily, the obstacle generates a reflected wave.

This behavior does no depend on a type of the inhomogeneous obstacle (dissipation, dispersion, or both) or on the form of distribution of inhomogeneity density. The number and parameters of resulting solitons vary, but the scenario stays invariable.

It is possible to predict the number, amplitudes and velocities o a wave that left the inhomogeneity obstacle using the comparative decay of the KdV conservation laws; some rough estimations are exemplified in subsection \ref{a priory}. A similar method may be applied to predict an evolution of an arbitrary initial compact datum for the KdV; details will be published soon.

\section*{Conclusion}

The transformation of initial soliton for the KdV equation with non-constant dissipation and/ordispersion was studied both numerically and analytically. In any such situation the transformation follows a definite pattern. So the results may be of a practical use. A form of a transformed wave, its reflection and  refraction coefficients may be   predicted. Thus the possibility of control of solitary impulses arises.

The figures in this paper were generated numerically using Maple PDETools package. The  mode of operation uses the default Euler method, which is a centered implicit scheme, and  can be used to find solutions to PDEs that are first order in time, and arbitrary order in space, with no mixed partial derivatives.

 Detailed algorithm for estimations of the refraction and reflection coefficients, based on the comparative decay of the selected KdV conservation laws will be published elsewhere.

\subsection*{Acknolegement}

This work was partially supported by the Russian Basic Research Foundation grant 18-29-10013.

\end{document}